\documentstyle[preprint,aps,prd,eqsecnum,amstex]{revtex}

\newcommand{\trv}{_{\perp}}
\newcommand{\gsim}{{_{\textstyle >} \atop ^{\textstyle \sim}}}
\newcommand{\lsim}{{_{\textstyle <} \atop ^{\textstyle \sim}}}
\newcommand{\Usla}{U\kern-1.6ex/}
\newcommand{\psla}{p\kern-1.0ex/}
\newcommand{\qsla}{q\kern-1.1ex/}
\newcommand{\esla}{\epsilon\kern-1.0ex/}
\newcommand{\etal}{{\it et al.~}}
\newcommand{\LQCD}{\Lambda_{\rm{QCD}}}
\newcommand{\Ha}[1]{{\rm H}_{#1}^{(1)}}

\newcommand{\BK}[1]{{\rm K}_{#1}}

\newcommand{\MM}{&}

\newcommand{\mpsi}{M_{\psi}}
\newcommand{\jp}{J/\psi}
\newcommand{\PR}[2]{Phys.~Rev.~#1#2}
\newcommand{\PRep}[1]{Phys.~Rep.~#1}
\newcommand{\PRL}[1]{Phys.~Rev.~Lett.~#1}
\newcommand{\PLB}[1]{Phys.~Lett.~B#1}
\newcommand{\NP}[2]{Nucl.~Phys.~#1#2}
\newcommand{\ZP}[2]{Z.~Phys.~#1#2} 

\newcommand{\ibid}[1]{{\it ibid.}~#1}
\begin{document}
\draft
\makeatletter
\date{\today}
\preprint{${\mbox{WU B 95-35}_{} \hfill} \atop {\mbox{hep-ph/9503289}}$}
\title{MODELLING THE NUCLEON WAVE FUNCTION \\
       FROM SOFT AND HARD PROCESSES}
\author{J. BOLZ \footnote{E-mail address:
                          bolz@wpts0.physik.uni-wuppertal.de}$\!$  
                \thanks{  Supported by the Deutsche
                          Forschungsgemeinschaft}
  and
        P. KROLL\footnote{E-mail address:
                          kroll@wpts0.physik.uni-wuppertal.de}$\!$
       }
\address{Fachbereich Physik, Universit\"at Wuppertal     \\
         D-42097 Wuppertal, Germany
        }
\maketitle
\newpage
%
%
\begin{abstract}
Current light-cone wave functions for the nucleon are unsatisfactory
since they are in conflict with the data of the nucleon's Dirac
form factor at large momentum transfer. Therefore, we attempt a
determination of a new wave function respecting theoretical ideas on
its parameterization and satisfying the following constraints: It
should provide a soft Feynman contribution to the proton's form factor
in agreement with data; it should be consistent with current
parameterizations of the valence quark distribution functions and
lastly it should provide an acceptable value for the $\jp \to N \bar
N$ decay width. The latter process is calculated within the modified
perturbative approach to hard exclusive reactions. A simultaneous fit
to the three sets of data leads to a wave function whose 
$x$-dependent part, the distribution amplitude, shows the same type of
asymmetry as those distribution amplitudes constrained by QCD sum rules. 
The asymmetry is however much more moderate as in those amplitudes. 
Our distribution amplitude resembles the asymptotic one in shape but 
the position of the maximum is somewhat shifted.
\end{abstract}
\narrowtext
\newpage 
%
%
\section{INTRODUCTION}
There is general agreement that the conventional hard scattering approach 
(see \cite{BrL80} and references cited therein) in which the collinear
approximation is used, gives the correct description of
electromagnetic form factors and perhaps of other exclusive
processes in the limit of asymptotically high momentum transfers. This
framework relies upon the factorization of hadronic amplitudes in
perturbative, short-distance dominated hard scattering amplitudes
and process-independent soft distribution amplitudes (DA).
In order to challenge arguments against the applicability of the hard
scattering approach in experimentally accessible regions of momentum 
transfer \cite{IsLlS} a modification of this scheme has been proposed
by Botts, Li and Sterman \cite{BoSLi} in which the transverse hadronic
structure is retained and gluonic radiative corrections in form of a
Sudakov factor are incorporated. This more refined treatment of
exclusive observables allows to calculate the genuinely perturbative
contribution self-consistently in the sense that the bulk of
the perturbative contribution is accumulated in regions of reasonably
small values of the strong coupling constant $\alpha_S$. Nevertheless
it turns out that the perturbative contributions to exclusive
observables are in general too small with perhaps a few exceptions. In
particular in the case of the nucleon's Dirac form factor it is shown
in Ref.~\cite{wubo} and is further confirmed in Ref.~\cite{Bol95} that
the perturbative contribution to it is indeed much smaller than the
experimental data. For a plausible value of the mean transverse
momentum inside the nucleon the perturbative contribution to the
proton form factor amounts to less than about 10 \% of the
experimental value \cite{wubo,Bol95}. \\
Ostensible agreement between data and calculations carried through
within the conventional hard scattering approach under use of
end-point concentrated DAs like that one proposed by  Chernyak \etal~(COZ)
\cite{COZ89a}, can be traced back to large contributions from the soft
end-point regions where one of the momentum fractions $x_i$ tends to
zero. In these regions gluon momenta become small and hence the use of
perturbation theory is unjustified \cite{IsLlS}. \\
The smallness of the perturbative contribution is, as we believe, not a
debacle. On the contrary it seems to be fully consistent within the entire
approach. We remind the reader of the contribution from the
overlap of the initial and final state (soft) nucleon wave
functions. This additional contribution, customarily termed the Feynman
contribution, is usually neglected but it is indeed large, as was
shown in Ref.~\cite{IsLlS} for a number of examples and as we are
going to demonstrate for a wide class of wave functions. For 
the end-point concentrated wave functions like those based on the COZ 
DA \cite{COZ89a}, the Feynman contributions even
exceed the experimental data \cite{pff} on the Dirac form factor of
the nucleon, $F_1^N$, by large amounts. This parallels observations
made on the pion's electromagnetic form factor recently (see, for
instance, \cite{JKR94}) according to which end-point concentrated 
pion wave functions are clearly disfavoured. \\
The purpose of the present paper is the construction of the nucleon's
(valence Fock state) wave function. In accord with the findings in
\cite{wubo,Bol95} we demand this wave function to provide a Feynman
contribution that completely controls the Dirac form
factor at momentum transfers around $10\,{\rm GeV}^2$. Admittedly,
this requirement does not suffice to determine the wave function,
further constraints are required. So we use the available information
on the parton distribution functions \cite{GRV94} to which the
nucleon's wave function is also related. As a third constraint we
employ the decay reaction $J/\Psi \to N\bar{N}$. This process is
expected to be dominated by perturbative contributions. We are going
to calculate the decay width for it within the modified perturbative
approach of Ref.~\cite{BoSLi} in contrast to previous analyses of the
$J/\Psi$ decay \cite{BrL81,COZ89b,BeS92,DTB85}. Employing a
parameterization of the wave function that complies with theoretical
ideas \cite{BHL83,ChZ95}, we determine the few (actually two)
parameters of the wave function from a combined fit to the data of the
three reactions just mentioned. We emphasize that we do not aim at a
perfect fit to the data, a number of theoretical uncertainties and 
approximations inherent in our approach would render such an attempt
meaningless. The purpose of our analysis is rather to demonstrate the
existence of a reasonable wave function from which the prominent
features of the data can be reproduced.\\  
The paper is organized as follows: In Sect.~II we briefly recapitulate
a few properties of the nucleon's light-cone wave function and we
introduce our ansatz for it. Sects.~III, IV and V are
devoted to the discussions of the Feynman contributions to $F_1^N$,
the parton distribution functions and the decay $\jp \to N \bar N$,
respectively. In Sect.~VI we will present a new wave function
satisfying the constraints discussed in Sects.~III, IV and V. Finally,
Sect.~VII contains our conclusions. An appendix includes a derivation
of the nucleonic $\jp$ width within the modified perturbative
approach. 
%
%
%
\section{THE NUCLEON WAVE FUNCTION}
Similarly to Sotiropoulos and Sterman \cite{Sot} we write the valence
Fock state of a proton with positive helicity as (the plane waves are
omitted for convenience)
\begin{eqnarray}
  \label{state}
  |\,P,+ \rangle\;
=
  \frac{1}{\sqrt{3!}} \varepsilon _{a_{1}a_{2}a_{3}}
  \int
  [{\rm d}x]
  [{\rm d}^{2}{\bf k_{\perp}}]
  \Bigl\{  & \phantom{} &
\!\!\!\!\!\!
         \Psi _{123}\,{\cal M}_{+-+}^{a_{1}a_{2}a_{3}} +
         \Psi _{213}\,{\cal M}_{-++}^{a_{1}a_{2}a_{3}}
\nonumber \\
& - &
         \Bigl(\Psi _{132}\, + \,
         \Psi _{231}\Bigr){\cal M}_{++-}^{a_{1}a_{2}a_{3}}
  \Bigr\}
\end{eqnarray}
where we assume the proton to be moving rapidly in the
3-direction. Hence the ratio of transverse to longitudinal momenta of
the quarks is small and one may still use a spinor basis on the
light cone. A neutron state is obtained by the replacement $u
\leftrightarrow d$. The integration measures are defined by
\begin{equation}
  \label{mass}
  \hspace{-1cm}
  [{\rm d}x] = \prod_{i=1}^3 {\rm d}x_i\,\delta(1-\sum_i x_i) \qquad
   [{\rm d}^2{\bf k\trv}] \equiv \frac{1}{(16\pi^3)^{2}}\,\prod_{i=1}^3 
     {\rm d}^2{\bf k_{\perp}}_i \, \delta^{(2)}(\sum_i {\bf k_{\perp}}_i) \:.
  \hspace{-1cm}
\end{equation}
The quark $i$ is characterized by the usual fraction $x_i$ of the
nucleon's momentum it carries, by its transverse momentum 
${\bf k_{\perp}}_i$ with respect to the nucleon's momentum as well as 
by its helicity and color. A three-quark state is then given by
\begin{equation}
  \label{quark}
  \hspace{-1cm}
  {\cal M}_{\lambda_1 \lambda_2 \lambda_3 }^{a_1 a_2 a_3} =
  \frac{1}{\sqrt{x_1 x_2 x_3}} \,
    |\,u_{a_1}; x_1, {\bf k_{\perp}}_1, \lambda_1\rangle \,
    |\,u_{a_2}; x_2, {\bf k_{\perp}}_2, \lambda_2\rangle \,
    |\,d_{a_3}; x_3, {\bf k_{\perp}}_3, \lambda_3\rangle.
  \hspace{-1cm}
\end{equation}
The quark states are normalized as follows
\begin{equation}
  \label{normq}
  \hspace{-1cm}
  \langle q_{a^{\prime}_i}; x^{\prime}_i, {\bf k_{\perp}}^{\prime}_i,
  \lambda^{\prime}_i \,|\, 
    q_{a_i}; x_i, {\bf k_{\perp}}_i, \lambda_i\rangle =
            2 x_i (2\pi)^3 \delta_{a^{\prime}_i a_i}
            \delta_{\lambda^{\prime}_i \lambda_i} \delta(x^{\prime}_i -x_i)
            \delta({\bf k^{\prime}_{\perp}}_i - {\bf k_{\perp}}_i)\:.
  \hspace{-1cm}
\end{equation}
Since the 3-component of the orbital angular momentum $L_z$ is assumed
to be zero the quark helicities sum up to the nucleon's helicity. 
As has been demonstrated explicitly in Ref.~\cite{Dzi88} (\ref{state})
is the most general ansatz for the $L_z\!=\!0$ projection of the
three-quark nucleon wave function: From the permutation symmetry
between the two $u$ quarks and from the requirement that the three
quarks have to be coupled in an isospin $1/2$ state it follows that
there is only one independent scalar wave function%
\footnote{In \cite{Dzi88} it has been shown that the entire
  nucleon state (including the $L_z\!\neq\!0$ projections) is
  described by three independent functions.}  
which, for convenience, we write as \cite{wubo} 
\begin{equation}
  \hspace{-2.5cm}
  \Psi _{123}(x,{\bf k_{\perp}})
\equiv
\Psi(x_1,x_2,x_3;{\bf k_{\perp}}_1,{\bf k_{\perp}}_2,{\bf k_{\perp}}_3)
=
  \frac{1}{8\sqrt{3!}}\,
  f_{N}(\mu_F)
  \phi_{123}(x,\mu _{F})\,
  \Omega (x,{\bf k_{\perp}})\:. \hspace{-1cm}
\label{Psiansatz}
\end{equation}
$f_N(\mu_F)$ plays the r\^ole of the nucleon wave function at the
origin of the configuration space and the factorization scale is 
denoted by $\mu_F$. $\phi_{ijk}(x,\mu_F) \equiv
\phi(x_i,x_j,x_k,\mu_F)$ is the nucleon DA conventionally normalized 
to unity
\begin{equation}
  \int [{\rm d}x] \phi_{123}(x, \mu_F) = 1\,. 
  \label{normDA}
\end{equation}
The DA is commonly expanded in a series of eigenfunctions $\tilde
\phi^n_{123}(x)$ of the evolution kernel being linear combinations of
Appell polynomials (see Ref.~\cite{BrL80}%
\footnote{Note that the $\tilde \phi_{123}^2(x)$ used here differs
  from that of Ref.~\cite{BrL80} by an overall sign.}%
)
\begin{equation}
  \phi_{123}(x,\mu_F) = \phi_{\rm AS}(x) \left[1 +
    \sum_{n=1}^{\infty} B_n(\mu_F)\,\tilde \phi^n_{123}(x) \right]
  \label{DAentw}
\end{equation} 
where $\phi_{\rm AS}(x) \equiv 120\,x_1 x_2 x_3$ is the
asymptotic (AS) DA \cite{BrL80}. Evolution is incorporated by the
scale dependences of $f_N$ and the expansion coefficients $B_n$ : 
\begin{equation}
  \hspace{-17mm}
  f_N(\mu_F)  =  f_N(\mu_0)\,\left(
    \frac{\ln(\mu_0/\LQCD)}{\ln(\mu_F/\LQCD)} \right)^{2/3\beta_0}
     \hspace{-8mm}, 
  \hspace{1cm}
  B_n(\mu_F)  =  B_n(\mu_0)\,\left(
    \frac{\ln(\mu_0/\LQCD)}{\ln(\mu_F/\LQCD)} \right)^{\tilde
    \gamma_n/\beta_0} \hspace{-8mm} 
  \hspace{5mm}
  \label{BnFNevol}
\end{equation}
where $\beta_0 \equiv 11 - 2/3\,n_f$ and $\mu_0$ denotes the scale of
reference customarily chosen to be 1 GeV. The exponents  $\tilde
\gamma_n$ are the reduced anomalous dimensions. Because they are
positive fractional numbers increasing with $n$ \cite{Pes79}, higher
order terms in (\ref{DAentw}) are gradually suppressed. \\   
Nucleon DAs are frequently utilized in applications of the
perturbative approach which are constrained by moments of the DA
\begin{equation}
  \phi^{(n_1 n_2 n_3)}(\mu_0) = \int [{\rm d}x]\,x_1^{n_1} x_2^{n_2}
  x_3^{n_3} \phi_{123}(x,\mu_0) \,
  \label{moments}
\end{equation}
evaluated by means of QCD sum rules \cite{COZ89a}. The few moments
known only suffice to determine the first five expansion coefficients 
$B_n$. However, since the moments are burdened by errors the $B_n$,
and hence the DA, are not fixed uniquely. In Ref.~\cite{BeS93}, for
instance, a set of 45 model DAs has been constructed where each DA
respects the moments of Ref.~\cite{COZ89a} and is strongly end-point
concentrated and asymmetric in the $x_i$. It is shown in
Ref.~\cite{wubo}, as we already mentioned in the introduction, that the
perturbative contributions to the nucleon form factor evaluated with
these DAs  are too small in comparison with the data \cite{pff}. As
we are going to discuss subsequently the DAs constructed by Bergmann
and Stefanis \cite{BeS93} also show serious deficiencies in other
applications. The value of $f_N$ is also determined from QCD sum rules
\cite{COZ89a}: $(5.0 \pm 0.3) \cdot 10^{-3}$ GeV$^2$. This value is
to be used in conjunction with the COZ DA \cite{COZ89a} and the DAs of
Ref.~\cite{BeS93}. \\ 
The transverse momentum dependence of the wave function is contained
in the function $\Omega$ which is normalized according to 
\begin{equation}
  \int[{\rm d}^2 {\bf k\trv}] \, \Omega(x,{\bf k\trv}) = 1 \,.
  \label{Omeganorm}
\end{equation}
Throughout we use a simple symmetric Gaussian parameterization for the
$k\trv$-dependence    
\begin{equation}
  \Omega(x,{\bf k_{\perp}})
=
  (16\pi ^{2})^{2}
  \frac{a^{4}}{x_{1}x_{2}x_{3}}
  \exp
     \left [
            -a^{2} \sum_{i=1}^{3}k_{\perp i}^{2}/x_{i}
     \right ]\:,
\label{BLHMOmega}
\end{equation}
which resembles the harmonic oscillator wave function proposed in
\cite{BHL83}. The ansatz (\ref{BLHMOmega}) keeps our model simple
and appears to be reasonable for a nucleon wave function $\Psi_{123}$
which is dominantly symmetric. With the ansatz (\ref{DAentw}),
(\ref{BLHMOmega}) antisymmetric or mixed symmetric contributions may
only appear through the DA. \\ 
For reasons which will become clear subsequently we only
need the soft part of the wave function, i.e.~the full wave function
with its perturbative tail removed from it \cite{BrL80}. 
The Gaussian $k\trv$-dependence is conform with the
behaviour of a soft wave function; the power-like decreasing
perturbative tail is removed. The ansatz (\ref{BLHMOmega}) is also
supported by recent work of Chibisov and Zhitnitsky \cite{ChZ95} who
showed that, on rather general grounds, $\Omega$ depends on $x_i$ and
${\bf k_{\perp}}_i$ solely in the combination $k_{\perp i}^2/x_i$ and that
$\Omega$ falls off like a Gaussian at large $k\trv$.%
\footnote{The kinematical transverse momentum of the partons is not
  the same object as $k\trv$ defined through moments as in
  \cite{ChZ95}. However, we will assume that both are one and the same
  variable. This assumption corresponds to summing up soft gluon
  corrections, i.e.~higher twist contributions.}
Eq.~(\ref{BLHMOmega}) is the simplest way to comply with these
requirements. We remark that integrating $\Omega$ in
Eq.~(\ref{Omeganorm}) to infinity instead of to a cut-off scale of
order $Q$ introduces only a small negligible error into the
calculation. \\  
According to (\ref{Psiansatz}) a wave function is defined by a certain
DA combined with the Gaussian (\ref{BLHMOmega}) and $f_N$. 
For the COZ wave function which we will subsequently confront to data
for the purpose of comparison, the transverse size parameter $a$,
controlling the width of the wave function in $k\trv$ space, is fixed
by requiring a certain value of either the root mean square (rms)
transverse momentum or the probability of the valence Fock state. We
will utilize the COZ wave function for three cases: $P_{3q} = 1$ ($a =
0.99$ GeV$^{-1}$, $\langle k\trv^2 \rangle^{1/2} = 272$ MeV),
$\langle k\trv^2 \rangle^{1/2} = 450$ MeV ($a = 0.60$ GeV$^{-1}$,
$P_{3q} = 0.13$) and 600 MeV ($a = 0.45$ GeV$^{-1}$, $P_{3q} =
0.04$). 
%
%
%
\section{THE FEYNMAN CONTRIBUTION}
The Dirac form factor of the nucleon can be expressed in terms of
overlaps of initial and final Fock state wave functions
\cite{DrY70,Wes70}. This is an exact representation of the form factor
provided a sum over all Fock states is implied and the full Fock state
wave functions are used. However, one can identify the overlaps of the
hard large $k\trv$ tails of the wave functions with the perturbative
contributions \cite{BrL80}. Since this contribution is small
\cite{wubo} the form factor is dominated by the overlaps of the soft
parts of the wave functions. The physical picture behind the overlap
representation is that one single quark is scattered by the virtual
photon with the remaining constituents following the struck quark as
spectators. The various overlap integrals are dominated by those
configurations where the struck quark carries almost the entire
momentum of the nucleon. Obviously, with an increasing number of
partons sharing the nucleon's momentum it becomes less likely that one
parton carries the full momentum of the nucleon. Therefore, higher
Fock state contributions are strongly suppressed at large momentum
transfer. The valence Fock state provides the most important soft
contribution (this is termed the Feynman contribution) in the region
of momentum transfer around 10 GeV$^2$. Of course, as required by the
consistency of the entire picture, the perturbative contribution will
take control in the limit $Q \to \infty$; the Feynman contribution
is suppressed by powers of $Q$ relative to the perturbative
contribution.  \\
%
%
According to Drell and Yan \cite{DrY70} we calculate the nucleon
matrix elements of the electromagnetic current in a frame where the
incoming nucleon is rapidly moving in the 3-direction (infinite
momentum frame). 
To leading order in the nucleon's momentum $P$ we
find from (\ref{state}), combined with the ansatz (\ref{Psiansatz}) for
the wave function, the following expression for the Feynman
contribution to the Dirac form factor
\begin{eqnarray}
  F_1^N{}_{\!soft}(Q^2) 
  \MM = \MM \left( \frac{f_N}{8 \sqrt{6}} \right)^2\,
      \sum_{j=1}^3 \,e_j\,  
      \int [{\rm d}x] \, \left[ \phi_{123}^2(x) + \phi_{213}^2(x)  
                      + (\phi_{132}(x)+\phi_{231}(x))^2 \right] \nonumber \\
  \MM \times \MM  \int [{\rm d}^2{\bf k\trv}] \,
      \Omega(x,{\bf k_{\perp}}_j + (1\!-\!x_j) {\bf q},
               {\bf k_{\perp}}_i - x_i {\bf q})\,
      \Omega(x,{\bf k\trv}) \:.
  \label{F1psi}
\end{eqnarray}      
This result is obtained from the matrix elements of the so-called good
current components ($\mu=0,3$). For $P\to\infty$ these matrix elements
are only fed by such configurations for which all constituents of the
nucleon move along the same direction as the nucleon (up to finite
transverse momenta), i.e. $0\leq x_i \leq 1$ for all constituents
\cite{DLY}. Matrix elements of the bad current components ($\mu=1,2$)
have to be treated with precaution; they are less reliable. Indeed,
these matrix elements are suppressed by $1/P$ as opposed to those of
the good current components and suffer from many approximations made,
e.g.~off-shell effects, helicity or spin rotations and so
on. Therefore, we also refrain from calculating the Pauli form factor,
$F_2$, along the same lines as $F_1$ since it is controlled by such
non-leading (with respect to the momentum $P$) contributions. 
To be more specific, $F_2$ can only be calculated if $L_z \neq 0$
components of the nucleon wave function are also included.  
We also note that the expression (\ref{F1psi}) 
somewhat differs from overlap expressions given by Isgur and
Llewellyn-Smith \cite{IsLlS} who start from current matrix elements in
a Breit frame and boost them to the infinite momentum frame. In this
procedure spin rotations have to be considered which, in a
model dependent way, generate $L_z\!\neq\!0$ components in the nucleon
wave function. Still the numerical results obtained in
Ref.~\cite{IsLlS} are very similar to our ones. \\ 
For a completely symmetric wave function the expression (\ref{F1psi})
is proportional to the sum of the quark charges and thus exactly
vanishes in the case of the neutron form factor $F_1^n$.%
\footnote{The neutron form factor is zero for any wave function at
      zero momentum transfer.}
This observation already precludes the AS wave function. For this reason
we refrain from showing results on $F_1^p{}_{\!soft}$ evaluated with
the AS wave function. The integrations in (\ref{F1psi}) can easily be 
performed analytically for wave functions of the type we consider 
consisting of the Gaussian (\ref{BLHMOmega}) and a DA for which the expansion
(\ref{DAentw}) is truncated at some finite $n$. For such wave
functions the Feynman contribution falls off proportional to $Q^{-8}$ for $Q^2
\to \infty$.  \\    
In Fig.~\ref{fig_feynm} we show the Feynman contribution to the proton
form factor evaluated with the COZ wave function%
\footnote{As the external $Q^2$ is the only scale present in the
  process it defines at the same time a natural evolution scale for
  the DA.}
in comparison with experimental data \cite{pff}.
It can be seen that the data is dramatically exceeded by the
Feynman contribution. This result is independent on the value of the
transverse size parameter used; an increase of the $k\trv$-width only
shifts the position of the maximum value of the Feynman contribution
to higher $Q^2$ without considerably reducing its magnitude. The
asymptotic $Q^{-8}$ behaviour does not set in before $Q^2 \approx 100$
GeV$^2$ (since the expansion of (\ref{F1psi}) into a series of $1/Q^2$
powers converges slowly). Similarly large Feynman contributions are
obtained from the COZ wave function in the case of the neutron. \\
The COZ wave function is representative for all strongly end-point
concentrated DAs. We checked that all wave functions constructed from
the Bergmann-Stefanis set of DAs \cite{BeS93} provide similar
unphysically strong Feynman contributions as the COZ wave function
\cite{Bol95}. As was found in \cite{IsLlS,CG87} an $x$-independent
Gaussian instead of our ansatz (\ref{BLHMOmega}), also leads to
large Feynman contributions to $F_1$ for the COZ DA. These findings
give rise to severe objections against DAs constrained by the QCD sum
rule moments of Ref.~\cite{COZ89a}. \\
The Feynman contributions are also subject to Sudakov corrections. An
estimate of these corrections on the basis of the Sudakov factor as
derived by Botts and Sterman \cite{BoSLi} (see (\ref{sudexp})) reveals
that the size of the Feynman contributions is somewhat reduced by
it. The suppression is stronger for the end-point concentrated wave
functions (but the Feynman contributions still exceed the form factor
data dramatically) than for the AS wave function or similar
ones. Since this suppression can be compensated for by readjusting the
wave function parameters we refrain from taking into account the
Sudakov corrections to the Feynman contributions.
%
%
%
\section{VALENCE QUARK DISTRIBUTION FUNCTIONS}

Deep inelastic lepton-nucleon scattering provides
information on the parton distributions inside nucleons. As is discussed in
\cite{BHL83} the parton distribution functions $q^N(x)$ representing
the number of partons of type $q$ with momentum fraction $x$ inside
the nucleon, are determined by the Fock state wave functions. 
%
Each Fock state contributes through the modulus squared of
its wave function, integrated over transverse momenta up to $Q$ and
over all fractions except those pertaining to partons of type
$q$. Obviously, the valence Fock state wave function only feeds the
valence quark distribution functions, $u^N_V(x)$ and $d_V^N(x)$. Since
each Fock state contributes positively the following inequalities hold 
\begin{eqnarray}
         u^p_V(x)  
     \MM \ge 2\MM\int [{\rm d}x] [{\rm d}^2{\bf k\trv}] \,
       \left[\Psi_{123}^2 + \Psi_{213}^2 + 
       (\Psi_{132} + \Psi_{231})^2\right]\,\delta(x-x_1) \,, 
     \nonumber \\ 
         d^p_V(x) 
     \MM \ge \MM  \int [{\rm d}x] [{\rm d}^2{\bf k\trv}] \,
       \left[\Psi_{123}^2 + \Psi_{213}^2 + (\Psi_{132} +
         \Psi_{231})^2 \right] \, \delta(x-x_3)  \,.
  \label{udp}
\end{eqnarray}
In (\ref{udp}) the full wave function enters. For reasons obvious from
the preceding discussions, we evaluate the valence Fock state 
contributions to $u_V^p$ and $d_V^p$ from our soft wave functions. 
On the strength of our experience with the nucleon form factor we
expect the neglect of the
perturbative tails to be admissible. The soft wave functions defined
by (\ref{Psiansatz}), (\ref{DAentw}) -- with the expansion truncated
at some finite order -- and (\ref{BLHMOmega}) leads to the behaviour
$x q_V^N(x) \sim (1-x)^3$ for $x \to 1$ in fair agreement with the
structure function data. This property is related to the asymptotic
behaviour of the Feynman contribution $F_1^N{}_{\!soft} \sim Q^{-8}$.
The interrelation between a $(1-x)^p$ behaviour of the distribution
functions for $x \to 1$ and a $(1/Q)^{p+1}$ behaviour of the Feynman
contribution at large $Q^2$ proposed by Drell and Yan \cite{DrY70} is
only obtained for wave functions factorizing in $x$ and $k\trv$
(i.e.~for $\Omega$ not depending on the $x_i$). Asymptotically, if the
perturbative contribution dominates the form factor, the Drell-Yan
interrelation also holds for wave functions of the type we consider. \\
Another interesting property of (\ref{udp}) is that any symmetric wave
function, as e.g.~the AS one, provides a value of 2 for the ratio
$u_V^p(x)/d_V^p(x)$. This is to be contrasted with the value of 
5 (at $x \approx 0.6$) for that ratio to be seen in current
parameterizations of the distribution functions \cite{GRV94}. \\
In Fig.~\ref{fig_QDF} we compare the valence Fock state contributions,
evaluated with the COZ wave function, to $u_V^p$ and $d_V^p$
with the Gl\"uck-Reya-Vogt parameterizations \cite{GRV94} at a
scale of 1 GeV$^2$ in the large $x$ region. For $x$ smaller than 0.6
our results well respect the inequalities (\ref{udp}). For $x$ larger
than about 0.6, on the other hand, wave functions normalized to large
probabilities (or equivalently providing small values of the mean
transverse momentum) violate the inequalities (\ref{udp}). For the COZ
wave function consistency with the Gl\"uck-Reya-Vogt parameterizations
can only be achieved if $P_{3q}$ is less than about 1\% which for our
wave function parameterization would correspond to unrealistically large
mean transverse momenta ( $\gsim 1$ GeV). Similar results are found
for the other COZ-like wave functions constructed from the
Bergmann-Stefanis set of DAs \cite{BeS93}. This observation is to be
considered as another serious failure of the COZ-like wave
functions. It should be mentioned that Sch\"afer
\etal~\cite{SMD89} found similar results for the valence quark
distribution functions or structure functions, respectively.  
%
%
%
\section{$\jp$ DECAY INTO NUCLEON-ANTINUCLEON}
The decay width of the process $\jp \to N \bar N$ provides a third
constraint on the nucleon wave function. By reason of the spin and the
parity of the $\jp$ meson and of color neutralization the massive charm
quarks the $\jp$ is composed of, dominantly annihilate through three
gluons which in turn create the light quark pairs necessary for the
formation of nucleon and antinucleon (see
Fig.~\ref{fig_jpsigraph}). There are hints that both the $c \bar c$
annihilation and the conversion of the gluons into $N \bar N$ pairs,
are under control of perturbative QCD. If, for instance, the decays of
charmonium states into light hadrons are viewed as decays into two or
three gluons the widths can be estimated perturbatively.
With acceptable values of $\alpha_S$ one obtains reasonable agreement
with experiment \cite{MP95} although in the case of the $\jp$ that
value appears to be a bit small with regard to the relevant scale 
provided by the charm quark mass (see the discussion in
\cite{CF94}). Another hint that perturbative QCD is at work in $\jp$
decays is provided by the angular distribution of $N \bar N$ pairs
produced via $e^+ e^- \to \jp \to N \bar N$. From the data of the DM2 
collaboration \cite{DM2} one estimates the fraction of $N \bar N$
pairs with equal helicities to amount to $(10 \pm 3)\%$ of the total
number of pairs. This is what is to be expected if the process is
dominated by perturbative QCD: Each of the virtual gluons produces a
light, almost massless quark and a corresponding antiquark with
opposite helicities. Since our nucleon wave function does not embody
any non-zero orbital angular momentum component the quark helicities
sum up to the nucleon's helicity. Hence, nucleon and antinucleon are 
dominantly produced with opposite helicities. This fact is an example
of the well-known helicity conservation rule for light hadrons
\cite{BrL80,BrL81}. The small amount of $N \bar N$ pairs with the
wrong helicity combination observed experimentally, while indicating
the presence of some soft contributions, can be tolerated. One should,
however, be aware of these contributions when theoretical results for 
the $\jp \to N \bar N$ decay are compared with experiment. \\
It is, however, fair to mention that there are several difficulties
with the perturbative calculation of exclusive charmonium decays,
e.g.~the relatively large branching ratios of $\psi'(2s) \to p \bar
p$, $\eta_c \to p \bar p$ and $\jp \to \pi \rho$. We
will nevertheless calculate the $\jp \to N \bar N$ decay width within
a perturbative approach. It will turn out and this may be regarded as
another argument in favour  of the prominent r\^{o}le of
perturbative QCD in charmonium decays, that the same
wave function that, on the one hand, leads to a very small perturbative
contribution to the nucleon form factor (and simultaneously to a large
Feynman contribution), provides, on the other hand, a reasonably large
value for the $\jp \to N \bar N$ decay width. \\
We start the calculation of the $\jp \to N \bar N$ decay width from an
invariant decomposition of the decay helicity amplitudes (with an
appropriately chosen spin quantization axis of the $J/\Psi$) 
\begin{equation}
  {\cal M}_{\lambda_1 \lambda_2 \lambda} = \bar u(p_1,\lambda_1)
  \left[ {\cal B}(\mpsi^2) \, \gamma_{\mu} + {\cal C}(\mpsi^2)\,
    \frac{(p_1 - p_2)_{\mu}}{2\,m_N} \right] v(p_2,\lambda_2) \, 
  \epsilon^{\mu}(\lambda) \, ,   
  \label{covzerl}
\end{equation}
where $p_1$ ($p_2$) and $\lambda_1$ ($\lambda_2$) are the momentum and
the helicity of the nucleon (antinucleon) respectively. $u$ and $v$
denote the nucleon spinors and $\epsilon$ the polarization vector of
the $\jp$. In the perturbative approach hadronic helicity conservation
forces the invariant function ${\cal C}$ to vanish. The $\jp \to N
\bar N$ decay width, therefore, takes the simple form (up to
corrections of order $(m_N/\mpsi)^4$ where $\mpsi$ is the $\jp$ mass)
\begin{equation}
   \Gamma(\jp \to N \bar N)  = \frac{\mpsi}{12\,\pi}\,|\,{\cal
     B}|^2\,. 
   \label{GamB}
\end{equation}
There is a small, almost negligible contribution to
the invariant function ${\cal B}$ from the $c \bar c$ annihilation
mediated by two gluons and a photon. The $gg\gamma$ contribution to
${\cal B}$ being proportional to the $ggg$ contribution, amounts to about 1\%
of the latter. In the following it is understood that the $gg\gamma$
contribution is absorbed into the $ggg$ contribution. There is also a 
small electromagnetic contribution to the invariant function ${\cal  B}$ 
from the $c \bar c$ annihilations through a virtual photon. This contribution
involves the time-like Dirac and Pauli form factor of the nucleon 
\begin{equation}
  {\cal B}_{\rm em} =  \frac{8\,\pi \alpha}{3\,\mpsi}\,f_{\psi}\,
    \left( F_1^N(s\!=\!\mpsi^2) + \kappa_N F_2^N(s\!=\!\mpsi^2)
    \right) \, .
\label{BCem}
\end{equation}
The factor multiplying the form factors is the invariant amplitude for
the $\jp$ decay into a lepton pair ($m_l = 0$). This factor includes
the $\jp$ decay constant $f_{\psi}$ (being related to the
configuration space wave function at the origin) which can be
determined from the experimental value of the leptonic decay widths 
\cite{pdg94}. One finds $f_{\psi} = 409 \pm 14$ MeV from a leading
order calculation. The $\alpha_S$ corrections to $f_{\Psi}$ are known
to be large \cite{BKKG}, leading to an increase of $f_{\Psi}$ by about
20\%. However, there is no assurance that the $\alpha_S^2$ corrections
are not also large and perhaps cancelling partly the $\alpha_S$
correction. Therefore we use the leading order value of $f_{\Psi}$
being aware of this eventual source of uncertainty in our final result. 
Using $F_1^p +\kappa_p F_2^p = G_M^p = 2.5 \pm 0.4$ GeV$^4 / \mpsi^4$ in
agreement with the E760 data \cite{E760} and $F_2^p = 0$, we estimate 
the electromagnetic contribution ${\cal B}_{\rm em}$ to the $\jp \to p
\bar p$ invariant decay amplitude ${\cal B}$ to amount to $ (14 \pm
4)\%$ of the experimental value. \\  
Several QCD studies of the decay $\jp \to 3 g \to N \bar N$ have appeared
in the past \cite{BrL81,COZ89b,BeS92,DTB85}. A point to criticize in
these studies which relied on the conventional hard scattering
approach, is the treatment of the strong coupling constant $\alpha_S$. 
Since, on the average, the virtuality of the intermediate
gluons is roughly 1 GeV$^2$ one would expect $\alpha_S$ to be
of the order of 0.4 to 0.5 rather than 0.2 to 0.3 which is usually
chosen \cite{COZ89b,BeS92,DTB85}. Since $\alpha_S$ enters to the sixth
power into the expression for the width a variation of $\alpha_S$
from, say, 0.3 to 0.45 would lead to a change by a factor of 11 for
the width. Thus, a large uncertainty is hidden in these calculations
preventing any severe test of the DA utilized. \\
In constrast to previous work \cite{BrL81,COZ89b,BeS92,DTB85} we will
not use the collinear approximation but rather use the modified 
perturbative approach of Sterman \etal~\cite{BoSLi} in which
transverse degrees of freedom are retained and Sudakov suppression,
comprising those gluonic radiative corrections not included in the
evolution of the wave function, are taken into account. The calculation
of the three-gluon contribution to the invariant function ${\cal B}$ is
presented in some detail in Appendix A.  \\
An important advantage of the modified perturbative approach is that 
the strong coupling constant can to be used in one-loop
approximation; its singularity, to be reached in the end-point regions
$x_i \to 0$, is compensated by the Sudakov factor. Hence, there is
no uncertainty in its use. This is to be contrasted with the conventional
perturbative approach where either $\alpha_S$ is evaluated at an
$x$-independent renormalization scale typically chosen to be of the
order of $\mpsi^2$ or at scales like (\ref{longscaljpsi}). In the
latter case for which, in contrast with the first case, large logs from
higher orders of perturbation theory are avoided, $\alpha_S$ has to be
frozen in at a certain value (typically 0.5) in order to avoid
uncompensated $\alpha_S$ singularities in the end-point regions. The
modified perturbative approach possesses another interesting feature:
the soft end-point regions are strongly suppressed. Therefore, the
bulk of the perturbative contribution comes from regions where the
internal quarks and gluons are far off-shell (order of $\mpsi^2$). In
contrast to the nucleon form factor the $\jp \to N \bar N$ amplitude
is not end-point sensitive.%
\footnote{This is due to the fact that in the collinear approximation
  the propagator denominators of the hard scattering amplitude vanish
  only linearly in the end-point regions whereas e.g.~in the case of
  the nucleon form factor they vanish quadratically for some of
  the $x_i$. }   
The suppression of the end-point regions does not, therefore, lead
to a substantial reduction of the $\jp \to N \bar N$ amplitude. \\  
As in Sects.~III and IV and before we turn to the determination of a
new nucleon wave function, we test the AS and the COZ wave
functions. Evaluating the three-gluon contribution ${\cal B}_{3g}$ 
from (\ref{B3gMHSP}) and leaving aside the inconsistencies with the 
form factor and the valence quark distribution functions we find 
acceptable values for the decay width ($\LQCD = 220$ MeV): Using the 
AS wave function ($f_N = 6.64 \cdot 10^{-3}$GeV$^2$, 
$a = 0.75$ GeV$^{-1}$) we find $\Gamma_{3g}(\jp \to p \bar p) = 0.09$ keV 
and for the COZ wave function ($a = 0.6$ GeV$^{-1}$) 0.23 keV. Smaller
(higher) values for the width are obtained if the transverse size 
parameter $a$ is decreased (increased). For comparison the experimental 
value for the $\jp \to p \bar p$ decay width is $0.188 \pm 0.014$ keV
\cite{pdg94}. Similar results for the width are obtained with the wave
functions constructed from the Bergmann-Stefanis DAs \cite{BeS93}. The
only exception is the Gari-Stefanis DA \cite{GS86} which provides a 
very small value for $\Gamma_{3g}$ of the order of a few eV. \\
%
%
%
%
%
\section{A MODEL FOR THE NUCLEON WAVE FUNCTION}
We have demonstated that the COZ-like wave functions as well as the AS
one do not describe the three processes discussed in the preceding
sections in a satisfactory manner. For this reason we will now try to
determine a new wave function from a fit to the proton form factor
data \cite{pff} (using (\ref{F1psi})), the valence quark distribution
functions of Gl\"uck \etal~\cite{GRV94} (using (\ref{udp})) as well as
the $\jp \to p \bar p$ decay width \cite{pdg94} (using
(\ref{B3gMHSP})). We start from the ansatz (\ref{state}), (\ref{Psiansatz}) 
and assume the Gaussian $k\trv$-dependence (\ref{BLHMOmega}) again. An
important question is, how many terms one has to allow in the expansion
(\ref{DAentw}). It would be obvious to truncate the expansion at $n=5$
again. As explorative fits immediately reveal, five terms in (\ref{DAentw}) 
provide too much freedom. For several expansion coefficients, in particular 
for $B_5$, one always obtains very small values. The corresponding terms 
in (\ref{DAentw}) can be neglected without worsening the fits noticeably. 
Indeed it suffices to consider only the first order expansion terms, 
$\tilde \phi_{123}^1$ and $\tilde \phi_{123}^2$, and even that still 
implies unnecessary freedom in case both the coefficients, $B_1$ and $B_2$ 
are treated as free parameters. As it turns out ultimately, the simple DA 
\begin{equation}
  \phi_{123} (x) = \phi_{\rm AS}(x) \left[ 1 + \frac{3}{4} 
    \tilde \phi_{123}^1(x) + \frac{1}{4} \tilde \phi_{123}^2(x)
  \right] = \phi_{\rm AS}(x) \frac{1}{2} \left[ 1 + 3 x_1 \right] 
  \label{phiFIT}
\end{equation}
meets all requirements. The only free parameters left over in this
case, namely $f_N$ and $a$, are determined by a fit to the data for
the three processes mentioned above. The fit provides the following
values for the two parameters: $f_N = 6.64 \cdot 10^{-3}$ GeV$^2$, $a
= 0.75$ GeV$^{-1}$. The fitted wave function implies a value of 0.17 for
the probability of the valence Fock state and a value of 411 MeV for
the rms transverse momentum. Both values appear to be reasonable. 
We stress that a larger flexibility in the DA, i.e.~allowing for more
free parameters to be adjusted in the fit, does not improve the fit
substantially. We also remark that the DA is not uniquely determined
by the data. Another solution of similar quality exists for which the 
DA contains the expansion terms $\tilde 
\phi_{123}^2$, $\tilde \phi_{123}^3$ and $\tilde \phi_{123}^4$. 
Although the mathematical expression of that DA looks rather
complicated the DA itself is similar to (\ref{phiFIT}) in shape and
magnitude. \\
Our value for $f_N$ is about 30 \% larger than that obtained from QCD
sum rules \cite{COZ89a}. In lattice QCD, on the other hand, the
following values for $f_N$ are found: $(2.9 \pm 0.6) \cdot 10^{-3}$
GeV$^2$ \cite{MaS89} and $6.6 \cdot 10^{-3}$ GeV$^2$
\cite{Bow88}. Thus, within a factor of about 2 all values agree with
each other. With regards to the large systematic uncertainties in the
various approaches the spread of the $f_N$ values cannot be considered
as a contradiction.\\
The proton decay offers another check of our wave
function. Calculating from it the three-quark annihilation matrix
element of the proton (termed $\alpha$ in \cite{BEHS}) along the same
lines as in \cite{BEHS}, we find for it a value of $0.012\, {\rm  GeV}^3$ 
which is about three times smaller than the value quoted in
Ref.~\cite{BEHS}. Our value is almost identical to a recent result
from lattice QCD \cite{Bow88} and rather close to many other results
\cite{Lan86}. The source of the difference between our result and
that one of Brodsky et al.\cite{BEHS} chiefly lies in the fact that we
give up the idea of calculating the proton form factor
perturbatively. \\ 
The DA (\ref{phiFIT}), displayed in Fig.~\ref{fig_FitDA}, possesses
interesting features. It is much less asymmetric and less end-point
concentrated than the COZ-like DAs which exhibit three pronounced
maxima and regions where the DAs acquire negative values. Our DA
rather resembles the AS one in shape but with the position of the only
maximum shifted to $x_1 = 0.44$, $x_2 = x_3 = 0.28$. Thus, as the COZ
DA but to a lesser amount, our DA possesses the property that, on
the average, a $u$-quark in the proton carries a larger fraction of
the proton's momentum than the $d$-quark. Related to the shift of the 
maximum's position is the asymmetry in the first order moments of our DA:
\begin{equation}
  \langle x_1 \rangle_{\rm FIT} = \frac{8}{21}\,,    \qquad
  \langle x_2 \rangle_{\rm FIT} = \langle x_3 \rangle_{\rm FIT} =
  \frac{13}{42} \, ,
  \label{momFIT}
\end{equation}
which values are to be contrasted with the QCD sum rule values
\cite{COZ89a} 
\begin{equation}
  \langle x_1 \rangle_{\rm COZ} = 0.54 - 0.62 \,,   \qquad
  \langle x_2 \rangle_{\rm COZ} = 0.18 - 0.20 \,,   \qquad
  \langle x_3 \rangle_{\rm COZ} = 0.20 - 0.25 \,. 
  \label{momCOZ}
\end{equation}
The moments (\ref{momFIT}) are consistent with those obtained by
Martinelli and Sachrajda from lattice QCD \cite{MaS89}. For a
discussion of the various approximations made in the QCD sum rule
analysis of \cite{COZ89a,CZ84r} and their implications see
\cite{Dun96}. \\
The DA (\ref{phiFIT}) or, when combined with the totally symmetric
Gaussian (\ref{BLHMOmega}), the wave function, can be
decomposed into a symmetric part and a part of mixed symmetry under
permutations (we adopt the notations of Ref.~\cite{IsLlS})
\[
  \phi^{S}(x)       = -\sqrt{6}         \,  \phi_{AS}(x)\,, \;\;
  \phi^{\rho}(x)    = -\frac{3}{2}(x_1-x_2) \phi_{AS}(x)\,, \;\;
  \phi^{\lambda}(x) =  \frac{\sqrt{3}}{2}(1- 3 x_3) \phi_{AS}(x)\,.
\]
These functions are related to the $\{56\}$ and $\{70,L\!=\!0\}$
representations of the permutation group on three
objects. Representations with non-zero orbital angular momentum $L$ do
not contribute to (\ref{phiFIT}). The symmetric $\{56\}$ part is
dominant since the ratio of probabilities $P_{3q}^S/P_{3q}$ is
$28/29 = 0.9655$. The strength of the $\{70,L\!=\!0\}$ admixture is
about the same as found from equal-time wave function analysis
\cite{Isg78}. This observation provides some justification of the
symmetric ansatz (\ref{BLHMOmega}) for the $k\trv$-dependence \`a
posteriori. \\ 
The results for the valence quark distribution functions obtained from
the fit are shown in Fig.~\ref{fig_QDF}. The valence Fock state
contribution to $x d_V(x)$ comes out comparatively large leaving
hardly room for contributions from higher Fock states for $x \gsim
0.6$. The effect of the asymmetric part of our DA provided by the 
eigenfunctions $\tilde \phi_{123}^1$ and $\tilde \phi_{123}^2$ is
clearly visible in Fig.~\ref{fig_QDF}: It pushes up $u_V$ at large $x$
and diminishes $d_V$ at the same time, thus producing a ratio $u_V :
d_V$ of about 5:1.\\
At this point it is in order to draw the reader's attention to a
little difficulty: The evolution behaviour of our soft contributions
to the valence quark distribution functions does not exactly match
with that of the phenomenological distribution functions. This entails
violations of the inequalities (\ref{udp}) for $Q^2 \gg 1$ GeV$^2$. The
imperfect evolution behaviour appears as a consequence of several
approximations made in our approach. Thus, for instance, we consider
only the soft part of the wave function and yet extend the upper limit
of the ${\bf k_{\perp}}$ integration to infinity (numerically this is
of little importance for a Gaussian like (\ref{BLHMOmega})\,). We also
ignored a possible evolution of the transverse size parameter (see 
\cite{BHL83} where the pion case is discussed). With respect to
our objective of a more qualitative understanding of the nucleon's
form factors and distribution functions rather than a perfect
quantitative description we tolerate that minor drawback. \\
In Fig.~\ref{fig_FitF1N} we show the results for the Feynman
contributions to the proton and neutron form factors in comparison with the
data \cite{pff,nff}. While the data on $F_{1}^{p}$ is input to the fit the
results for the neutron form factor are predictions. It can be
seen that our wave function provides Feynman contributions which
exhibit a broad maximum near 15 GeV$^2$. The asymmetric part of the
DA (\ref{phiFIT}) is solely responsible for the neutron form factor (pushing it
down from zero to a negative value) and pushes up the Feynman
contribution to the proton form factor (see the difference between
the solid and dashed lines in Fig.~\ref{fig_FitF1N}). For $Q^2$ smaller
than about 8 GeV$^2$ the fit is somewhat below the data. We are
content with that result because, as we already mentioned, our goal is
not a perfect fit but rather to demonstrate the existence of a wave
function that provides a Feynman contribution to the nucleon's form
factor of the right magnitude and that is consistent with the
constraints from the other two processes we consider and thus
implicitly gives an explanation for the smallness of the perturbative
contribution to the nucleon's form factor. An improved fit in
agreement with the form factor data over a wide range of $Q^2$ can
likely be achieved with minor modifications of the Gaussian
(\ref{BLHMOmega}) at finite transverse momentum. Since such
modifications, an example of which has been given by Zhitnitsky
\cite{Zhi94} for the case of the pion wave function, require more free
parameters we persist in the simple Gaussian (\ref{phiFIT}). For  
$Q^2 \lsim 8$ GeV$^2$ the presence of higher Fock states is to be 
expected (remember $P_{3q} = 0.17$). The DA of a Fock state consisting
of $n_g$ gluons and $n_q$ quarks  contains to lowest order in the 
momentum fractions terms proportional to 
$x_1 x_2 ... x_{n_q} x_{n_q+1}^2 ... x_{n_q+n_g}^2$ as
is supported by power counting arguments given in \cite{CZ84r}. Using
this asymptotic form of a higher Fock state DA in combination with a 
Gaussian $k\trv$-dependence of the type (\ref{BLHMOmega}) one finds 
the Feynman contribution to fall off as $Q^{-4(n_q + 2 n_g  -1)}$. 
Hence, for this type of wave functions contributions from higher Fock 
states become strongly suppressed at large $Q^2$ but may be quite 
important below $\sim 8$ GeV$^2$. \\  
Our wave function provides the value   
\begin{equation}
  \Gamma_{3g}(\jp \rightarrow N \bar N) = 0.117 \mbox{ keV} 
  \label{GamFIT}
\end{equation}    
for the three gluon contribution to the nucleonic $\jp$ width. 
If comparing the result with experiment \cite{pdg94} one has to
be aware of the electromagnetic contribution and the spin effect.
The electromagnetic contribution to ${\cal B}$ (see
(\ref{BCem})) amounts to about 15 \% (see Sect.~V). Since, however, 
the phase of the experimental time-like form factor is unknown we are
not in the position to add ${\cal B}_{3g}$ and ${\cal B}_{\rm em}$ coherently
at present.%
\footnote{In a common perturbative analysis of both the proton
  form factor and the $J/\Psi$ decay, the relative phase between
  the two contributions would be fixed.}
Thus, we can only say that, at best, the electromagnetic
contribution may increase our prediction by 30 \%. Considering also
that $N \bar N$ pairs with equal helicity contribute about 10 \% to
the total width, we regard our result for the three-gluon contribution
as consistent with experiment. \\
Other sources of uncertainties in our calculation of the $\jp$ decay 
widths are introduced by the value of $f_{\Psi}$ chosen (see the
discussion in Sect.~5) and by $\LQCD$. 
Since $\alpha_S$ enters to the sixth power in $\Gamma_{3g}$ the result
(\ref{GamFIT}) is rather sensitive to the value of $\LQCD$
employed. A mild increase of $\LQCD$ by 10 \% enlarges $\Gamma_{3g}$
by about 30 \% without changing the results shown in
Figs.~\ref{fig_QDF} and~\ref{fig_FitF1N} noticably; only the evolution
behaviour of the nucleon wave function in the calculation of the
Feynman contribution is slightly modified. Thus, we conclude that 
our wave function (\ref{BLHMOmega}), (\ref{phiFIT})
provides a reasonably large three gluon contribution
to the $\jp \to p \bar p$ decay width. In contrast to previous
calculations of this width carried through in collinear approximation,
our average $\alpha_S$ (being 0.43) is consistent with the available
scale in the $\jp$ decay which is provided by the $c$-quark mass. We
also note that our perturbative calculation is self-consistent. The
bulk of the contribution, i.e.~more than 50 \%, is accumulated in
regions where $\alpha_S^3 < 0.47^3$. \\ 
At this point a remark concerning the $n \bar n$ decay channel
is in order. Since the three-gluon contribution is flavour-blind any
difference between the decay widths into $p \bar p$ and $n \bar n$
must be due to the electromagnetic contribution. From experiment it is
known that the widths for $\jp \to p \bar p$ and $\jp \to n \bar n$
agree within the experimental errors \cite{Ant93a} and that the
time-like form factors for the proton and the neutron, to which the
${\cal B}_{\rm em}$  are directly proportional, are  approximately
equal in modulus at $s = 5.4$ GeV$^2$ \cite{Ant93b}. Since both the
contributions, ${\cal B}_{3g}$ and ${\cal B}_{\rm em}$, are in
general complex numbers with non-trivial phases the only conclusion to
be drawn at present is that the relative phase between ${\cal B}_{3g}$ and
${\cal B}_{em}$ is the same (up to an eventual sign) for the proton
and the neutron channel. \\ 
The generalization to the decay $\Upsilon \to N \bar N$ is a
straightforward task. For bottomium systems the application of our
approach is even better justified than for charmonia because of
the larger gluon virtualities. Using a value of 710 MeV for the decay
constant $f_{\Upsilon}$ of the $\Upsilon$ meson which, in a similar
manner as $f_{\psi}$, is determined from the leptonic width of the
$\Upsilon$ \cite{pdg94}, we find $\Gamma_{3g}(\Upsilon \to N \bar N) =
1.3 \cdot 10^{-2}$ eV. The electromagnetic contribution to $\Upsilon
\to N \bar N$ decays is negligible as an estimate in analogy to that
one in the $J/\Psi$ case shows. Up to now there is only an
experimental upper limit of about 1 eV for this decay width
\cite{pdg94} which is safely met by our result. The
effective $\alpha_S$ is 0.26 in the $\Upsilon$ case.\\
At the end of this section we want to comment on other forms of the
nucleon's wave function to be found in the literature. For instance,
$x$-independent parameterizations of $\Omega$ are used sometimes
\cite{IsLlS,CG87}. Although such factorizing forms of the wave
functions are in conflict with rotational invariance and with the
arguments given in Ref.~\cite{ChZ95}, the numerical results, say, for
the Feynman contribution to the nucleon form factor obtained with
them do not differ much from our results. In other cases the wave
function is regarded as a constituent wave function. Consequently
the parameters are adjusted so that the wave function is normalized
to unity and that the charge radius of the nucleon is reproduced. 
Constituent wave functions also provide large Feynman contributions
which, however, decrease more rapidly with increasing momentum
transfer than our results obtained from a wave function describing an
object that is smaller than the nucleon. The introduction of form
factors for the constituent quarks mediating the transition to current
quarks, may improve the large momentum transfer behaviour of the form
factor (see, for instance, Ref.~\cite{CPSS}). Another possibility to
construct a wave function is to start from an equal time wave function
(e.~g.~that of a harmonic oscillator) in the nucleon's rest frame and 
transform it to the light cone under the assumption that a Melosh 
transform for free, non-interacting quarks can be applied (see, 
for instance, Ref.~\cite{Dz92}). While this is a very interesting approach
attempting to construct a unified picture of the non-relativistic
quark model and light-cone physics, the resulting wave function
presented in Ref.~\cite{Dz92} does not pass our tests against
data. \\
Occasionally DAs are used which possess a multiplicative factor
$\exp[-a^2 \sum_i m_q^2/x_i]$ (see, for instance,
Refs.~\cite{BHL83,Dz92}) where the parameter $m_q$ is to be
interpreted as a constituent quark mass. Although the infinite series
(\ref{DAentw}) can, in principle, accommodate such a mass exponential,
the truncated expansion may not reproduce it sufficiently accurate. It
may perhaps be better to consider the mass exponential explicitly. To
get an idea about the importance of that mass exponential we modify
the DA (\ref{phiFIT}) by it and repeat our calculations. It turns out
that the $\jp$ decay width as well as the distribution functions are
mildly affected by the mass exponential while the Feynman contribution
is reduced by about 20 \% in the momentum transfer region between 5
and 15 GeV$^2$ but decreases somewhat faster with increasing $Q^2$
than the Feynman contribution obtained from the original DA
(\ref{phiFIT}). The reduction around 10 GeV$^2$ can be compensated for
by a 10 \% increase of $f_N$. \\
Finally, one may think of additional powers of transverse
momenta multiplying the Gaussian (\ref{BLHMOmega}) (see, for instance, 
Ref.~\cite{Dz92}). Despite of this and many other possible
complications we stick to our simple ansatz (\ref{state}), (\ref{Psiansatz})
and (\ref{BLHMOmega}) because, as we have shown, it is flexible enough
to account for the available data with a sufficient degree of accuracy.  
%
%
\section{SUMMARY AND CONCLUSIONS}
In this paper we have investigated the soft light-cone wave function
of the nucleon. For the DAs we use the expansion in terms of the 
eigenfunctions of the evolution equation truncated at some finite
order. For the transverse momentum dependence of the wave function we
use a specific form which is supported by results obtained in
\cite{BHL83,ChZ95}. On the strength of rather general arguments the
$k_{\perp}$-dependence of the wave function appears in the form
$k_{\perp i}^2/x_i$ with a Gaussian fall-off at large $k_{\perp}$.
Studying soft Feynman contributions to the Dirac form factor
$F_1^N$ of the nucleon and valence Fock state contributions to
the quark distribution functions of the nucleon at large $x$ we found 
that wave functions constructed on the basis of QCD sum rules in
general lead to too large predictions. These results resemble similar
considerations in the case of the pion \cite{JKR94}.
In the pion case there is strong evidence that
the DA is close to the asymptotic one, whereas in the case 
at hand asymmetries between the proton and neutron form factor as well
as structure function data prevent the use of the asymptotic DA.\\
We take these observations as a motivation to model a new wave
function which consistently describes the valence Fock state
contributions to the quark distribution functions and the Dirac form
factor $F_1^N$ by its Feynman contribution. In addition we require
that it also leads to a proper prediction for the $\jp \to N \bar {N}$
decay  width within the modified perturbative approach. The
perturbative calculation is self-consistent in that case and, 
in contrast to the case of the nucleon form factor, 
the perturbative contribution is in fair agreement with the
experimental result on the decay width. It should be noted that there
is still some residual uncertainty in the perturbative contribution
from the imperfect know\-ledge of the strong coupling constant
$\alpha_S$ and the $\jp$ decay constant. At any rate we have been
able to rectify the treatment of $\alpha_S$ by avoiding some fixed
scale prescriptions in contrast to previous calculations. In our
calculation the effective value of $\alpha_S$ is 0.43 for our wave
function at $\LQCD = 220$ MeV instead of about 0.3 as in previous
calculations.\\ 
The wave function which we determine from the combined fit to the
three sets of data consists of a Gaussian $k_{\perp}$-dependence and a
very simple DA which bears resemblance to the asymptotic DA in shape
but with the position of the only maximum shifted somewhat. Like the
COZ-type DAs our DA possesses the interesting property that, on the
average, a $u$-quark in the proton carries a larger fraction of the
proton's momentum than the $d$-quark. Our wave function defined by
(\ref{state}), (\ref{Psiansatz}), (\ref{BLHMOmega}) and
(\ref{phiFIT}), has only two free parameters to be adjusted. With more
complicated wave functions, containing more free parameters, the fit
can certainly be improved. However, with regard to our aim of
demonstrating the existence of a soft nucleon wave function which
complies with theoretical ideas and from which the prominent features
of the data can be reproduced, we refrain from introducing such complications. 

\acknowledgements

We would like to thank N.G.~Stefanis for useful discussions and
S.J.~Brodsky for valuable comments.
\begin{appendix}
%
%
\section{PERTURBATIVE CALCULATION OF THE $\jp \to N \bar N$ DECAY
  AMPLITUDE} 
As in previous perturbative calculations
\cite{BrL81,COZ89b,BeS92,DTB85} the $\jp$ meson will be treated as a
non-relativistic $c \bar c$ system with $v^2/c^2$ corrections
neglected. According to \cite{Kue79} we write the $\jp$ state in a
covariant fashion
\begin{equation}
  |\,J/\psi; \,q,\lambda\,\rangle
  \,=\, \frac{\delta_{ab}}{\sqrt{3}}\,\frac{f_{\psi}}{2\,\sqrt{6}}\,
        \frac{(\qsla+\mpsi) \esla(\lambda)}{\sqrt{2}} \,,
  \label{Jpsistate}
\end{equation}
where $a$ and $b$ are color indices. 
Within the modified perturbative approach the three-gluon contribution
${\cal B}_{3g}$ to the $\jp$ decay into $N \bar N$ is of the form  
\begin{eqnarray}
  \MM\MM {\cal B}_{3g} =  \frac{f_{\psi}}{2\sqrt{6}}\,
     \int [{\rm d}x] [{\rm d}x'] \int\,\frac{{\rm d}^2{\bf b}_1}{(4\pi)^2}\,
     \frac{{\rm d}^2{\bf b}_3}{(4\pi)^2} \,
     \hat T_H(x,x',{\bf b})\,\exp[-S(x,x',\mpsi)] 
     \label{B3gMHSP} \\ 
  \MM\MM \times \, 
  \big[ \hat\Psi_{123}(x,{\bf b}) \hat\Psi_{123}(x',{\bf b})  
         + \frac{1}{2}
      \left(\hat\Psi_{123}(x ,{\bf b} ) +  \hat\Psi_{321}(x ,{\bf b} )
                                                               \right)\!
      \left(\hat\Psi_{123}(x',{\bf b}) +  \hat\Psi_{321}(x',{\bf b})
                                                               \right)
     \big] 
     \,.\nonumber
\end{eqnarray}
This convolution of wave functions and a hard scattering amplitude $T_H$
can formally be derived by using the methods described in detail by
Botts and Sterman \cite{BoSLi}. 
The ${\bf b}_i$, canonically conjugated to the transverse momenta ${\bf
k_{\perp}}_i$, are the quark separations in the transverse
configuration space. ${\bf b}_1$ and ${\bf b}_3$ correspond to the
locations of quarks 1 and 3 in the transverse plane relative to quark
2 and ${\bf b}_2 = {\bf b}_1 - {\bf b}_3$.  
$\hat\Psi_{ijk}$ represents the Fourier transform of the wave function
$\Psi_{ijk}$.  \\
The hard scattering amplitude, to be calculated from Feynman graphs of
the type shown in Fig.~\ref{fig_jpsigraph}, reads
\begin{equation}
  T_H(x,x',{\bf k\trv},{\bf k\trv'}) 
    =  \frac{5120\sqrt{6}/27\,\pi^3\,\mpsi^5\,(x_1 x'_3 + x_3 x'_1)}
       {[\tilde q_1^2+({\bf k_{\perp}}_1\!+\!{\bf k_{\perp}'}_1)^2]
       \,[\tilde q_3^2+({\bf k_{\perp}}_3\!+\!{\bf k_{\perp}'}_3)^2]}\,
       \prod_{i=1}^3  \frac{\alpha_S(t_i)} 
       {\tilde g_i^2-({\bf k_{\perp}}_i\!+\!{\bf k_{\perp}'}_i)^2
         +i\epsilon}
       \:, \hspace{0.5cm}
   \label{THjpsi}
\end{equation}
where 
\begin{equation}
  \tilde q_i^2 = [ x_i\,(1-x_i') + (1-x_i)\,x_i']\,\mpsi^2/2
      \, , \hspace{1cm}
  \tilde g_i^2 =  x_i x_i'\,\mpsi^2 \,. \hspace{0.5cm}
  \label{longscaljpsi}
\end{equation}
Note that $T_H$ depends on sums of transverse momenta, ${\bf K}_i
\equiv {\bf k_{\perp}}_i + {\bf k'_{\perp}}_i$, and because of the
constraint $\sum_i {\bf K}_i = {\bf 0}$
only two of the ${\bf K}_i$, say ${\bf K}_1$ and ${\bf K}_3$, are
independent. Hence, the Fourier transformed hard amplitude 
\begin{equation}
  \hspace{-1cm}
  \hat T_H(x,x',{\bf b}_1, {\bf b}_3) 
  =  \int 
     \frac{{\rm d}^2{\bf K}_1}{(2\pi)^2}\frac{{\rm d}^2{\bf K}_3}{(2\pi)^2}\,
     T_H(x,x',{\bf K})\,
     \exp [ -i\,{\bf K}_1 {\bf \cdot b}_1 
           - i\,{\bf K}_3 {\bf \cdot b}_3]\,, \hspace{-1cm}
  \label{THfour1}
\end{equation}
depends only on the vectorial distances ${\bf b}_1$ and ${\bf b}_3$. In
physical terms this means that the $N \bar N$ pairs emerge with 
identical transverse separation configurations from the decay because
each gluon produces a quark-antiquark pair at the same location in the
transverse configuration plane which thereafter do not interact. We
emphasize that in contrast to the case of the nucleon form factor
\cite{wubo} this circumstance is not due to approximations concerning
the $k\trv$-dependences of $T_H$ but instead a direct consequence of
the decay kinematics.\\     
Inserting (\ref{THjpsi}) into (\ref{THfour1}), one finds for the hard
scattering amplitude in ${\bf b}$ space
\begin{eqnarray}
  \hat T_H(x,x',{\bf b}_1, {\bf b}_3) 
  \MM = \MM
     -\frac{2560}{27} \,f_{\psi}\,\mpsi^5\,
     \frac{(x_1 x'_3 + x_3 x'_1)}
     {[\tilde q_1^2 + \tilde g_1^2]
      [\tilde q_3^2 + \tilde g_3^2]} \,\frac{1}{(2\pi)^3} \,
     \prod_{i=1}^3 (\pi\alpha_S(t_i))\,
     \int {\rm d}^2{\bf b}_0\,\hspace{1cm}
     \nonumber \\
  \MM \times \MM
     \left[ \frac{i\pi}{2} \Ha{0}(\tilde g_1\,|{\bf b}_1 + {\bf b}_0|\,)
     - \BK{0}(\tilde q_1\,|{\bf b}_1 + {\bf b}_0|\,) \right] \,
     \frac{i\pi}{2} \Ha{0}(\tilde g_2 b_0) \,
     \nonumber \\
  \MM \times \MM
     \left[ \frac{i\pi}{2} \Ha{0}(\tilde g_3\,|{\bf b}_3 + {\bf b}_0|\,)
     - \BK{0}(\tilde q_3\,|{\bf b}_3 + {\bf b}_0|\,) \right] \: . 
  \label{THfour2}
\end{eqnarray}
The auxiliary variable ${\bf b}_0$ in Eq.~(\ref{THfour2}) serves
as a Lagrange multiplier to the constraint $\sum {\bf K}_i = {\bf 0}$.  
Inserting (\ref{THfour2}) into the expression (\ref{B3gMHSP}) for the
invariant function ${\cal B}_{3g}$ we see that a nine dimensional
numerical integration is to be performed.%
\footnote{Taking into account relativistic corrections to the $\jp$
  wave function, i.e.~its transverse momentum dependence, one would
  have to perform a 14 dimensional numerical integration which is
  impossible with present day computers to a sufficient degree of
  accuracy.} 
Although this is a rather involved technical task it can be carried
through with sufficient accuracy if some care is put into it. 
Since the virtualities of the gluons are timelike, $\hat T_H$ includes
complex-valued Hankel functions $\Ha{0}$ which are related to the
usual modified Bessel functions $\BK{0}$, appearing for space-like
propagators, by analytic continuation in the momentum transfer
variable which in our case is the $\jp$ mass. Thus, ${\cal B}_{3g}$
has a non-trivial phase as for instance the time-like form factors
\cite{PiG95}. \\ 
The Sudakov factor $\exp[-S]$ in (\ref{B3gMHSP}) takes into
account those gluonic radiative corrections not accounted for in the
QCD evolution of the wave function as well as the renormalization group
transformation from the factorization scale $\mu_F$ to the
renormalization scales $t_i$ at which the hard amplitude $\hat T_H$ is
evaluated. The Sudakov exponent reads
\begin{equation}
  \hspace{-1cm}
  S(x,x',\mpsi) = \sum_{i=1}^3 \left[
    s(x_i,\tilde b,\mpsi) + s(x'_i,\tilde b,\mpsi) 
    + \frac{4}{\beta}\log\frac{\log(t_i \LQCD)}{\log(1/\tilde b_i
      \LQCD)} 
  \right]\:,\!
  \hspace{-1cm}
  \label{sudexp}
\end{equation}
where $\beta \equiv 11 - 2/3 n_f$. The function $s(\xi,\tilde
b,\mpsi)$, originally derived by Botts and Sterman \cite{BoSLi} and
later on slightly improved, can be found in Refs.~\cite{Bol95,DJK95}. 
$\mpsi$ appears in the function $s$ since it provides the large scale in the
process of interest. \\ 
The quantities $\tilde b_i$ are infrared cut-off parameters, naturally
related to, but not uniquely determined by the mutual separations of
the three quarks \cite{CoS81}. Following \cite{wubo} we chose $\tilde
b_i = \tilde b = \max\{b_1,b_2,b_3\}$. With this ``MAX'' prescription
the hard scattering amplitude is unencumbered by $\alpha_S$
singularities in the soft end-point regions. As a consequence of the
regularizing power of the ``MAX'' prescription, the perturbative
contribution saturates in the sense that the results become
insensitive to the inclusion of the soft regions. A saturation as
strong as possible is a prerequisite for the self-consistency of the
perturbative approach. 
The infrared cut-off $\tilde b$ marks the interface betweeen the
non-perturbative soft gluons, which are implicitly accounted for in
the nucleon wave function, and the contributions from soft gluons,
incorporated in a perturbative way in the Sudakov factor. Obviously,
the infrared cut-off serves at the same time as the gliding
factorization scale $\mu_F$ to be used in the evolution of the wave
function. \\
The renormalization scales $t_i$ are defined in analogy to the case of
electromagnetic form factors \cite{BoSLi,wubo} as the maximum scale of
either the longitudinal momentum or the inverse transverse separation
associated with each of the gluons
\begin{equation}
   t_1 = \max(\tilde q_1, \tilde g_1, 1/b_3) \,, \hspace{0.5cm}
   t_2 = \max(\tilde g_2, 1/b_2)             \,, \hspace{0.5cm}
   t_3 = \max(\tilde q_3, \tilde g_3, 1/b_1) \,.
   \label{tijpsi}
\end{equation}
The above assignment of $b$-scales is not compelling. Rearrangements
in the $b$-scales, however, induce only slight changes in the
numerical results.  
\end{appendix}
%
%

%
%
%
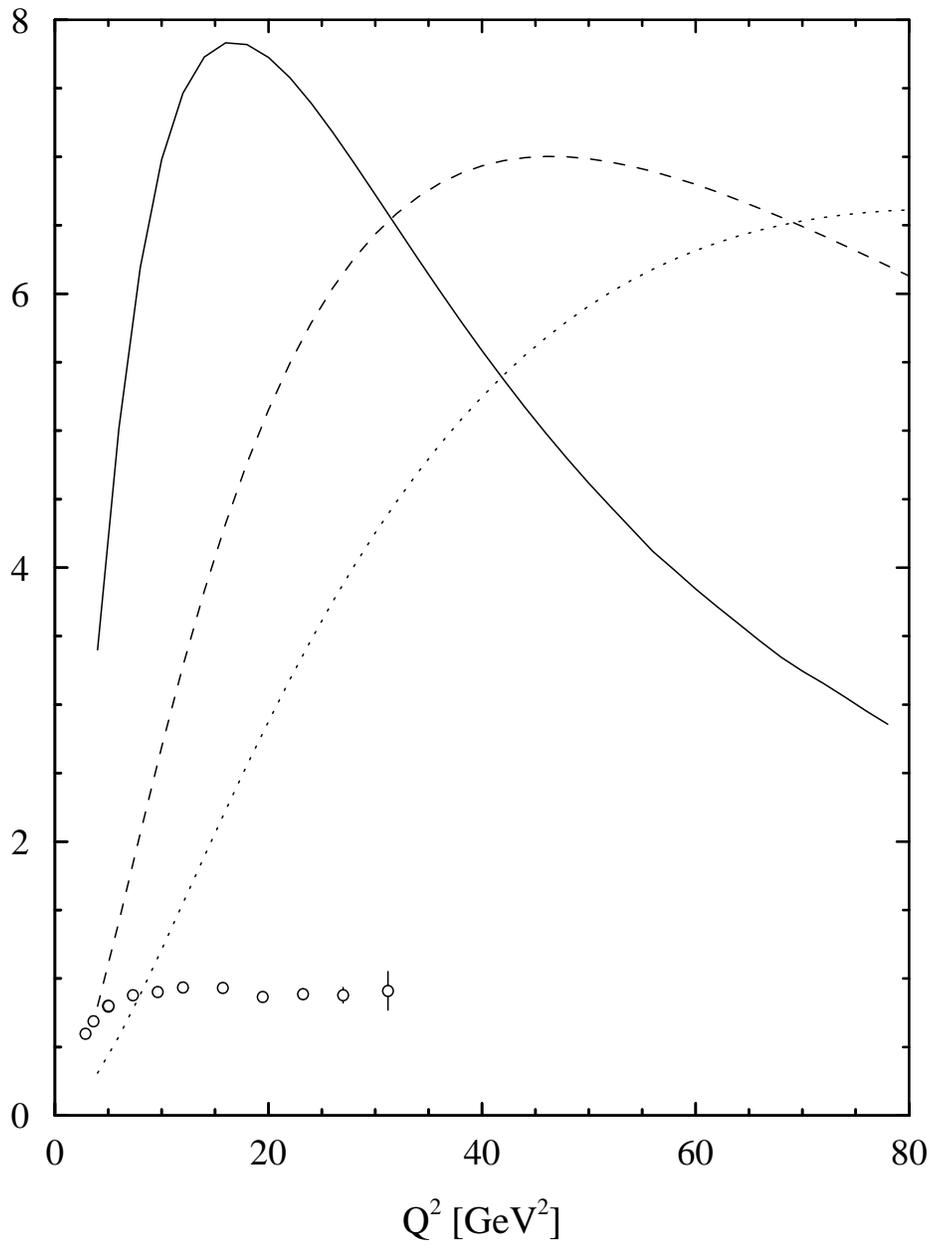
\begin{figure}
  \caption[dummy1]{
    Feynman contributions to Dirac form factor $F_1^p$ using the COZ
    wave function. The solid (dashed, dotted) line is evaluated with
    $a = 0.99$ (0.60, 0.45) GeV$^{-1}$. Experimental data ($\circ$)
    are taken from \cite{pff}.  
    \label{fig_feynm}}
\end{figure}
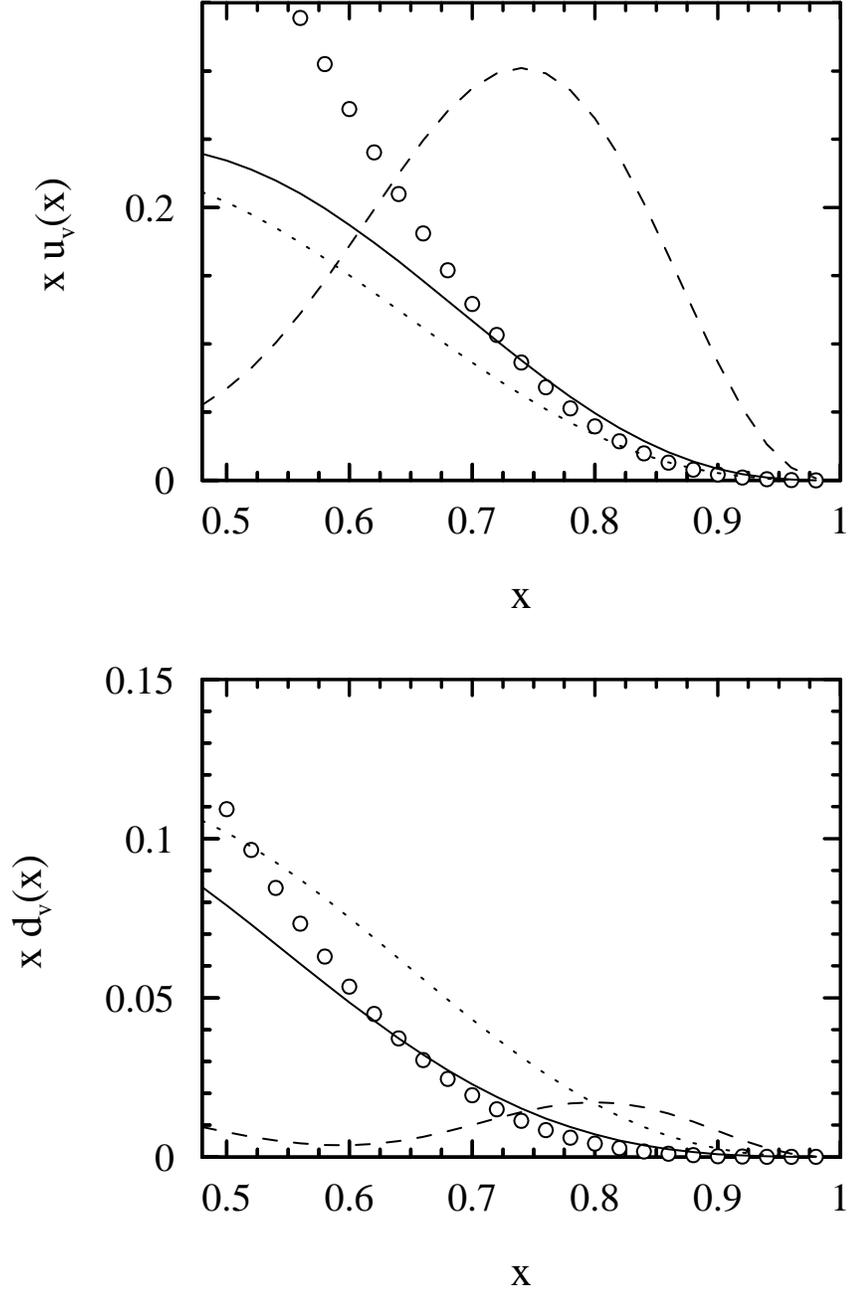
\begin{figure}
\caption[dummy2]
{ Valence Fock state contributions to the valence quark distribution
  functions of the proton at $Q^2 = 1$ GeV$^{2}$. The open circles
  represent the parameterization of Ref.~\cite{GRV94}. The solid and
  dashed lines represent the contributions of the valence Fock state
  using a wave function composed of the Gaussian (\ref{BLHMOmega})
  and either the DA (\ref{phiFIT}) or the COZ one ($a = 0.60$
  GeV$^{-1}$), respectively. The dotted line is obtained from the AS
  wave function with $f_N$ and $a$ as for the wave function
  (\ref{BLHMOmega}), (\ref{phiFIT}). 
  \label{fig_QDF}}
\end{figure}
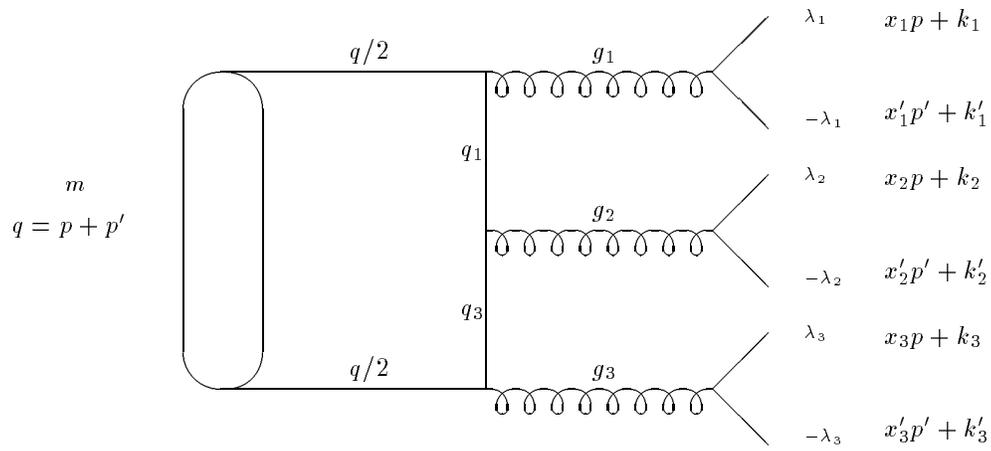
\begin{figure}
\caption[dummy3]{Decay graph $\jp \to 3g \to 3 q \bar q$. The momenta
  of the quarks are $x_i p+ k_i$ with $k_i = (0,0,{\bf k}_{\perp i})$,
  and those of the antiquarks are marked by a prime.
  \label{fig_jpsigraph}}
\end{figure}
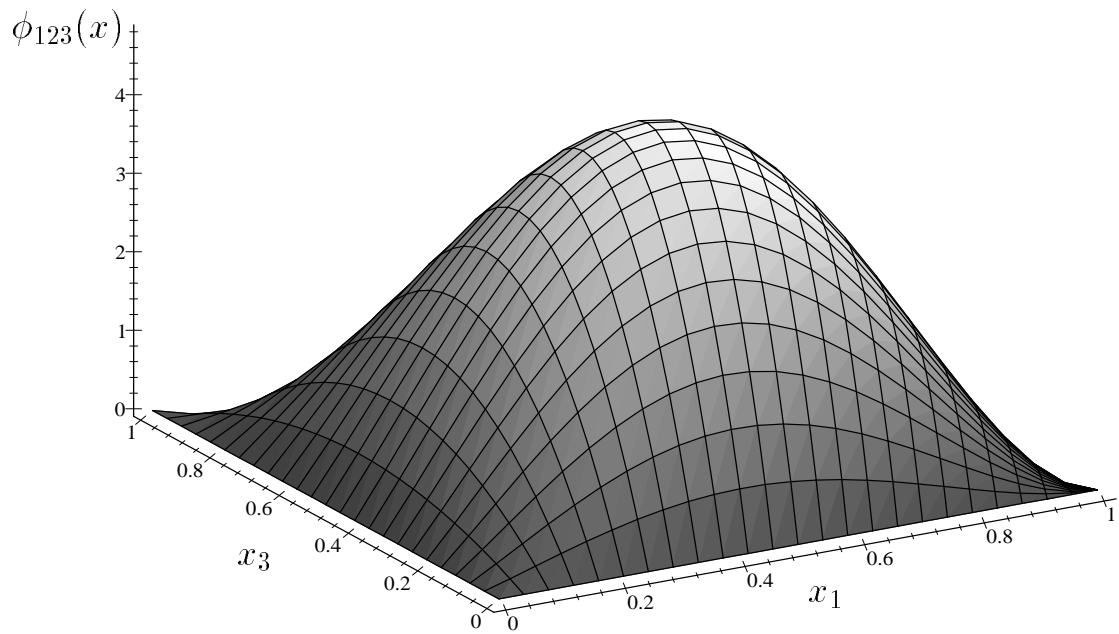
\begin{figure}
  \caption[dummy4]{The DA (\ref{phiFIT}) as a function of $x_1$ and $x_3$. 
  \label{fig_FitDA}}
\end{figure}
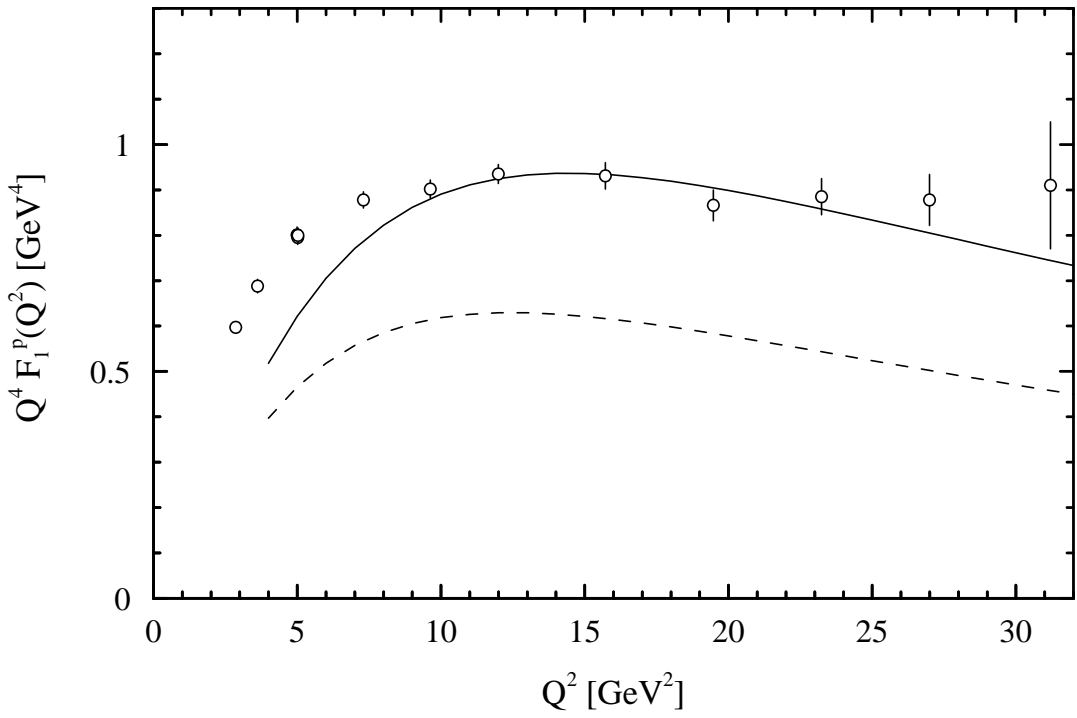
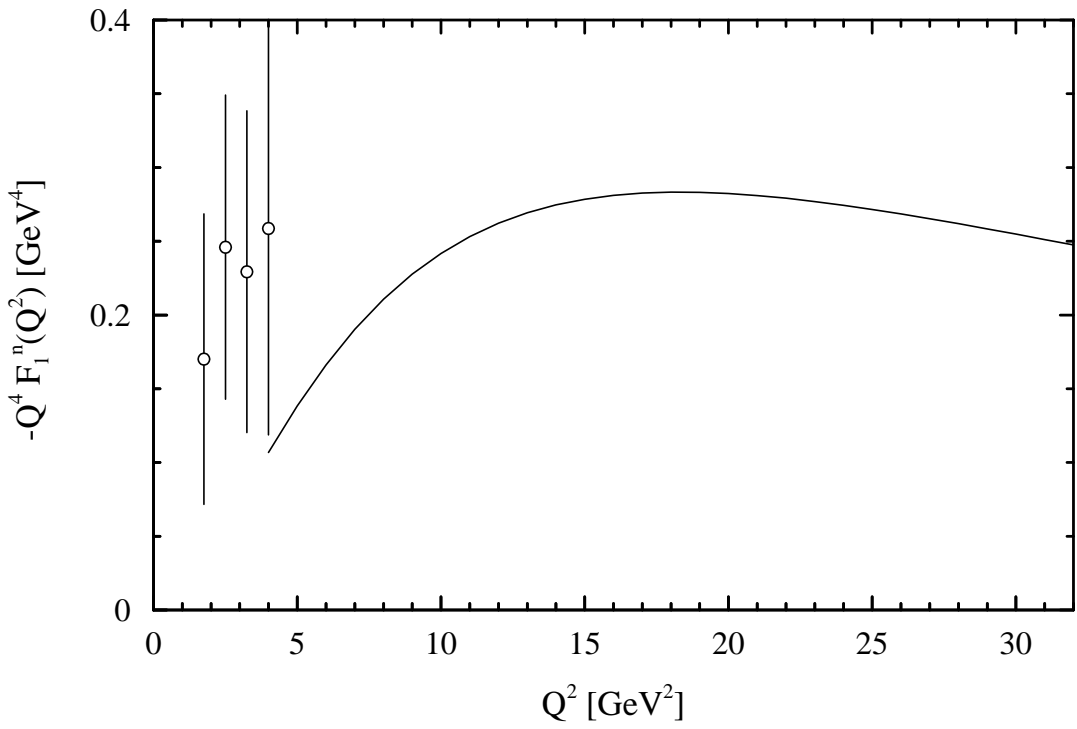
\begin{figure}
  \caption[dummy5]{Feynman contribution to the Dirac form factor of
    the proton (top) and the neutron (bottom) evaluated from the wave
    function (\ref{BLHMOmega}), (\ref{phiFIT}). The dashed line in 
    the upper figure is obtained from the AS wave function with $f_N$ 
    and $a$ as for the wave function (\ref{BLHMOmega}), (\ref{phiFIT}). 
    Data ($\circ$) are taken from Refs.~\cite{pff,nff}.   
  \label{fig_FitF1N}}
\end{figure}
\end{document}